\documentclass[letterpaper, 10 pt, conference]{ieeeconf}  

\IEEEoverridecommandlockouts                              

\usepackage{graphicx}
\usepackage{amsmath}
\usepackage{amssymb}
\usepackage{soul,color}
\usepackage{placeins}
\usepackage{tikz}
\usepackage{booktabs}
\usepackage[caption=false, font=footnotesize]{subfig}

\usetikzlibrary{arrows.meta,positioning}

\newcommand \auxdRelgz	{ \kappa_4 }

\newcommand \auxRelcz		{ \kappa_1 }

\newcommand \auxRelgz		{ \kappa_3  }

\newcommand \dRelgz   	{ \kappa_4 }
\newcommand \dzref		{ \dot z_\mathrm{d} }
\newcommand \ddzref		{ \ddot z_\mathrm{d} }
\newcommand{\dtau}{\mathrm{d}\tau}

\newcommand \ksp  { k_\mathrm{s} }

\newcommand \lambdasat	{ \kappa_2 }

\newcommand \param		{ p } 
\newcommand \npar		{ q } 
\newcommand \parnom {\param^*} 
\newcommand \parnor {\theta} 
\newcommand \parnornom {\parnor^*} 

\newcommand \Rel			{ \mathcal R }

\newcommand \Relcz		{ \kappa_1 }

\newcommand \Relgz		{ \kappa_3 }
\newcommand \senu		{S} 
\newcommand \tf			{ t_\mathrm{f} }

\newcommand \vcoil		{ u }
\newcommand \uff {\vcoil_{f{\!}f}}

\newcommand \vcap         {\\ \vspace{-0.84ex}}
\newcommand \vel			{ v }

\newcommand \zf			{ z_\mathrm{f} }
\newcommand \zref			{ z_\mathrm{d} }
\newcommand \zsp  { z_\mathrm{s} }

\let\originalleft\left
\let\originalright\right
\renewcommand{\left}{\mathopen{}\mathclose\bgroup\originalleft}
\renewcommand{\right}{\aftergroup\egroup\originalright}

\title{\LARGE \bf
Faster Run-to-Run Feedforward Control of Electromechanical\\Switching Devices: a Sensitivity-Based Approach*
}

\author{Edgar Ramirez-Laboreo, Eduardo Moya-Lasheras, and Eloy Serrano-Seco
    \thanks{* This work was supported in part via grants \mbox{PID2021-124137OB-I00}, TED2021-130224B-I00, and CPP2021-008938, funded by MCIN/AEI/ 10.13039/501100011033, by ERDF A way of making Europe, and by the European Union NextGenerationEU/PRTR, in part by the Government of Arag\'on - EU, under grant T45{\_}23R, in part by the ``Programa Investigo'' funded by the European Union - Next Generation EU, and in part by Fundaci\'on Ibercaja and the University of Zaragoza, via grant \mbox{JIUZ2023-IA-07}.}
    \thanks{The authors are with the Departamento de Informatica e Ingenieria de Sistemas (DIIS) and the Instituto de Investigacion en Ingenieria de Aragon (I3A), Universidad de Zaragoza, 50018 Zaragoza, Spain, {\tt\small \{ramirlab, emoya, eserranoseco\}@unizar.es}}%
}

\begin{document}

\maketitle
\thispagestyle{empty}
\pagestyle{empty}

\begin{abstract}
Electromechanical switching devices, such as solenoid valves, contactors, and relays, suffer from undesirable phenomena like clicking, mechanical wear, and contact bounce. Despite that, they are still widely used in industry due to their various economic and technical advantages. This has encouraged the development of controllers aimed at reducing the collisions that occur at the end of the switching operations. One of the most successful approaches has been the use of iterative techniques. However, these algorithms typically require a large number of operations to converge, which is definitely a clear drawback. This paper presents a strategy to improve the convergence rate of such controllers. Our proposal, which is based on the sensitivity of the control law with respect to the parameters, assumes that the performance of the system is more heavily affected by some parameters than others. Thus, by avoiding movements in the directions that have less impact, the search algorithm is expected to drive the system to near-optimal behaviors using fewer operations. Results obtained by simulation show significant improvement in the convergence rate of a state-of-the-art run-to-run feedforward controller, which demonstrates the high potential of the proposal.
\end{abstract}

\section{Introduction}

Solenoid valves are commonly used in hydraulic circuits~\cite{van2019design} or internal combustion engines~\cite{DiGaeta2015}, while electromechanical relays can be found in power circuits, medical devices, automotive applications and, in general, in virtually all industries~\cite{bojan2019design}. These are just a few examples of the widespread use of electromechanical switching devices. The open-loop dynamics of all of them is similar: when energized, a magnetic force attracts a moving component of the device, causing it to continuously increase its speed until the end of the stroke, where a violent impact occurs, generating acoustic noise and gradually wearing out the device. Despite this drawback, no alternatives have yet been found that can compete with the very low cost and numerous electrical and mechanical strengths of these devices. For this reason, research on modeling, analysis, and control of electromechanical switching devices is still being published.

In order to reduce the undesirable phenomena associated with impacts, several control schemes have been proposed that aim to reach the final position with zero velocity. This is commonly known as soft-landing control. Different approaches are reported in the literature: backstepping control~\cite{deschaux2019magnetic}, sliding-mode control~\cite{fang2019modeling}, extremum-seeking adaptive control~\cite{Benosman2015}, or iterative learning control~\cite{hoffmann2003iterative}, among others. Given the fast and nonlinear dynamics of these devices, any controller can benefit from model-based feedforward terms~\cite{braun2019flatness}, as these generally allow for improved response time and tracking accuracy. In extreme cases, where neither the position nor any other related variable can be measured for technical or economic reasons, the controller could even be based solely on the feedforward term and thus implemented in an open-loop fashion. There is however an obvious problem with this strategy: any difference between the dynamics of the real system and that of the dynamical model used to design the controller will result in a loss of performance. These differences may be due, e.g., to inconsistencies in the model equations or errors in the estimation of its parameters, or simply changes in the system dynamics due to wear or varying ambient conditions. 

One way to minimize the problems associated with feedforward controllers is to exploit the repetitive operation of these devices to update the controller parameters on a cycle-by-cycle basis. The only necessary condition for the application of iterative methods is the existence of measurable variables that allow, albeit indirectly, to evaluate the performance of the system in a given operation. For example, the performance of electromechanical relays has been successfully improved using run-to-run (R2R) algorithms based on measurements of the electrical contacts~\cite{ramirez2017new} or the acoustic noise generated when switching~\cite{serrano2022}. The key idea is to transform the problem into a black-box optimization. For that purpose, the input must be defined by a finite set of parameters and the available measurements must be fed into a cost function that evaluates the performance of the system. The relation between the parameters that define the input signal and the cost is obviously unknown, since it depends both on the system dynamics---which may not be perfectly modeled---and on the possible disturbances and randomness inherent to each operation. For that reason, it is advantageous to use direct search optimization methods to solve the problem, since these do not require information about the gradient and can even be used to optimize discontinuous cost functions. The so-called pattern search algorithms~\cite{lewis2000pattern} are of particular interest because of their very low computational cost. Bayesian optimization algorithms are also a suitable solution when more computational resources are available~\cite{moya2019novel}.

Although iterative algorithms have proven to be a successful solution for improving the performance of electromechanical devices, their main disadvantage is that they require a large number of switching operations until convergence is reached. To address this problem, in this paper we present two methods aimed at improving the convergence rate of this class of control algorithms, both based on the sensitivity of the control law with respect to the parameters. Although sensitivity is a well-established and widely used concept in areas such as system identifiability~\cite{miao2011identifiability} and parameter estimation~\cite{walter1997identification}, to the best of our knowledge, it has not yet been applied to improve iterative control algorithms. Specifically, we have applied this methodology to improve the convergence of a recently published adaptive R2R control algorithm for electromechanical relays~\cite{moya2023IFAC}.

The paper is organized as follows. Sections~\ref{sec:model} and~\ref{sec:R2R} present, respectively, the dynamical model used to design the controller and the R2R algorithm. Section~\ref{sec:theory} explains the sensitivity-based approach to improve convergence, including the two specific methods that we propose. Section~\ref{sec:simulation} contains simulation results that demonstrate the functionality of our two proposals. Finally, the conclusions are discussed in Section~\ref{sec:conclusions}.

\section{System dynamics} \label{sec:model}

In this section, we provide a brief description of the dynamics of the system to control. For a more detailed explanation, readers are referred to our works~\cite{serrano2022} and \cite{moya2023IFAC}.

The electromechanical switching devices under study are single-coil reluctance actuators, consisting of a fixed core wound with a current-carrying coil and a movable iron core. When current flows through the coil, the fixed core becomes magnetized, attracting the movable core. In addition to the magnetic force, the motion of this component is also affected by passive elastic forces, which can generally be modeled as ideal springs. The dynamics of the system can be described using a state-space model, with the position $z$, velocity $\vel$, and magnetic flux linkage $\lambda$ as state variables. The coil voltage, $u$, is the input. The corresponding state equations are
\begin{align}
    \dot z &= \vel, \label{eq:dynz} \\
    \dot \vel &= \frac{1}{m}\, \left( -k_\mathrm{s} \, \left(z-z_\mathrm{s}\right) - \frac{1}{2}\,\lambda^2 \, \frac{\partial \Rel}{\partial z} \right), \\
    \dot \lambda &= -R \, \lambda\, \Rel(z,\lambda) + u, \label{eq:dynlambda}
\end{align}
where $m$ is the moving mass, $k_\mathrm{s}$ is the spring stiffness, $z_s$ is the spring resting position, $R$ is the coil resistance, and $\mathcal R$ is an auxiliary function based on the magnetic reluctance concept. To account for the phenomena of magnetic saturation and flux fringing in the model, $\mathcal R$ is defined as
\begin{equation}\label{eq:Rel}
    \Rel(z,\lambda) =  \frac{\Relcz}{1-|\lambda|/\lambdasat} + \Relgz + \frac{\dRelgz \, z}{1 + \kappa_5\, z \, \log(\kappa_6/z)},
\end{equation}
where $\auxRelcz$, $\lambdasat$, $\auxRelgz$, $\auxdRelgz$, $\kappa_5$, and $\kappa_6$ are positive constants. Overall, the system dynamics depends on $q=9$ uncertain parameters, which can be grouped in the parameter vector $p$.
\begin{equation}\label{eq:param0}
        \param = \\ \left[
        \,\, k_\mathrm{s} \,\,\,\, z_\mathrm{s} \,\,\,\, m \,\,\,\, \auxRelcz \,\,\,\, \lambdasat \,\,\,\, \auxRelgz \,\,\,\, \auxdRelgz \,\,\,\, \kappa_5 \,\,\,\, \kappa_6 \,\, 
    \right]^\intercal
\end{equation}
Note that the resistance $R$ is treated independently as a parameter without uncertainty, as it can be precisely measured. 

The model \eqref{eq:dynz}--\eqref{eq:dynlambda} exhibits differential flatness when the position $z$ is regarded as its output. This property allows us to express the state variables and the input as functions of the flat output and its derivatives, as well as the parameter vector $\param$. In particular, the input $\vcoil$ can be expressed as
\begin{equation}\label{eq:u_flat}
	\vcoil = f_\vcoil\left(z, \dot z, \ddot z, \dddot z, p \right) =
	R\,\Rel(z,\lambda) \, \lambda + \dot\lambda,
\end{equation}
where $\lambda$ and $\dot \lambda$ are given by
\begin{align}
	\lambda = f_\lambda(z, \ddot z, p) = \sqrt{\frac{-2\,\big( k_\mathrm{s} \, \left(z-z_\mathrm{s}\right) + m \, \ddot z\big)}{\frac{\partial \Rel}{\partial z}}},\label{eq:phi_flat} \\
	\dot \lambda =  f'_\lambda\left(z,\dot z, \ddot z, \dddot{z}, \param\right) = \frac{ -\ksp\,\dot{z} - m \dddot{z} - \frac{1}{2}\,\lambda^2\,\frac{\partial^2\Rel}{\partial z^2}\, \dot z}{\lambda\,\frac{\partial \Rel}{\partial z}}.\label{eq:dphi_flat}
\end{align}

\section{Run-to-run feedforward control}
\label{sec:R2R}

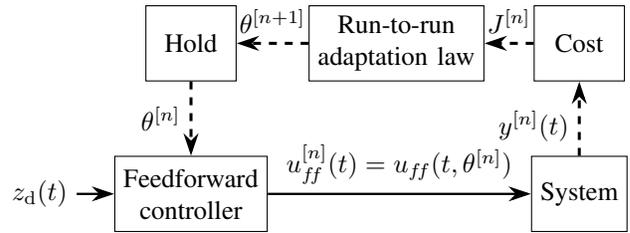
\begin{figure}[t]
	\def\sumoffset{1mm}
	\def\sumoffsetaux{1.5mm}
	\def\nodex{10mm}
	\def\nodey{13mm}
	\def\arrowsep{3mm}
	\def\lwt{0.3mm}
	\def\lwn{0.4mm}
	\def\lwd{0.5mm}
	\begin{tikzpicture}[
		node distance = \nodey and \nodex,
		box/.style = {draw, minimum height=10mm, minimum width=12mm, align=center},
		sum/.style = {circle, draw, node contents={}},
		>={Stealth[width=2mm,length=3mm]}
		]
		\node (ref) [] {$\zref(t)$};
		\node (ff) [box, right=of ref, xshift=-5mm] {Feedforward\\controller};
 		\coordinate[right=of ff,xshift=90] (c2);
		\node (plant) [box, right=of ff,xshift=25mm] {System};
		\coordinate[below=of plant.center, yshift=-0.5*\arrowsep] (c5);
		\coordinate[above=of plant.center] (c7);
		\coordinate[above=of c7] (c8);
		\coordinate[above=of c8] (c9);
		\coordinate[above=of ff.center] (c10);
		\node (hold) [box, above=of ff.center,yshift=20, anchor=center] {Hold};
		\node (cost) [box, above=of plant.center,yshift=20, anchor=center] {Cost};
		\node (opt) [box, right=of hold,xshift=32, anchor=center] {Run-to-run\\adaptation law};
		\draw[->,line width=\lwt] (ff) -- node[above] {$\uff^{[n]}(t)=\uff(t,\parnor^{[n]})$} (plant);
		\draw[dashed,->,line width=\lwn] (plant) -- node[left] {$y^{[n]}(t)$} (c7) -- (cost);
		\draw[dashed,->,line width=\lwn] (cost) -- node[above] {$J^{[n]}$} (opt);
		\draw[dashed,->,line width=\lwn] (opt) -- node[above] {$\parnor^{[n+1]}$} (hold);
		\draw[dashed,->,line width=\lwn] (hold) -- node[left] {$\parnor^{[n]}$} (ff);
		\draw[->,line width=\lwt] (ref) -- (ff);
	\end{tikzpicture}
    \caption{Control diagram. The superscript $[n]$ denotes the variables of the $n$th operation. The feedforward block computes $\uff$ from the parameter vector $\parnor$ and the desired trajectory $\zref$. The adaptation law updates $\parnor$ once per operation using the cost $J$, which is derived from the measurable output~$y$.}
	\label{fig:v_ctrl_diag}
\end{figure}

The R2R control strategy presented in~\cite{moya2023IFAC} and used as the basis for this work is schematized in Fig.~\ref{fig:v_ctrl_diag}. It has two main components: a feedforward controller, based on differential flatness theory, and an iterative adaptation law for the controller parameters.

\subsection{Feedforward control law}
The feedforward control law is based on the input function derived using the flatness property, as expressed in~\eqref{eq:u_flat}, which takes a position signal, its derivatives, and the model parameters as inputs. To use this expression as a feedforward control law, a desired position trajectory $\zref$ must firstly be designed. Considering that the objective is to achieve soft landing, $\zref$ can be designed as a $5$th-degree polynomial with the following boundary conditions:
\begin{equation}\label{eq:bounds}
    \begin{aligned}
	\zref(t_0) &= z_0, & \dzref(t_0) &= 0, & \ddzref(t_0) &= 0, \\
	\zref(\tf) &= \zf, & \dzref(\tf) &= 0, & \ddzref(\tf) &= 0,
    \end{aligned}
\end{equation}
where $t_0$ and $\tf$ are the desired initial and final times of the switching operation, and $z_0$ and $\zf$ are the desired initial and final positions, which correspond to the mechanical limits of the motion of the movable core.

Once the desired trajectory is set, the feedforward control law can be expressed as a function of a set of parameters with physical interpretation. For reasons related to the adaptation law, each parameter is calculated as the product of its constant nominal value and a dimensionless control parameter. That is, the feedforward control law, $\uff$, is defined as
\begin{equation}\label{eq:uff}
    \uff(t,\theta) = f_u\left(\zref(t), \dot z_{\mathrm{d}}(t), \ddot z_{\mathrm{d}}(t), \dddot z_{\!\!\mathrm{d}}(t), \parnom\! \odot \parnor \right),
\end{equation}
where $\parnor\in \mathbb{R}^q$ is the vector of control parameters, $\parnom\in \mathbb{R}^q$ is the nominal parameter vector, and $\odot$ denotes the element-wise---or Hadamard---product. 

\subsection{Adaptation law}
One of the drawbacks of feedforward controllers is their dependence on the accuracy of the system model and parameter identification. To mitigate this problem, a run-to-run adaptation law is incorporated. This law acts as an online black-box optimization algorithm that iteratively updates the parameter vector $\theta$ of the feedforward control law. The optimization objective is to minimize a cost function $J$ calculated from an output variable of the device, namely the absolute value of the impact velocity, denoted as $v_\mathrm{c}$.
\begin{equation}\label{eq:cost}
    J= \left\lvert {v_\mathrm{c}}\right\rvert
\end{equation}

In this context, the chosen optimization method is the pattern search algorithm~\cite{lewis2000pattern}, a derivative-free direct search numerical optimization method. It works by repeatedly applying a specific pattern in a multidimensional search space. Specifically, each evaluated point differs from the central point of the mesh---the best solution so far---in only one of the coordinates. The convergence rate is a key consideration, as each evaluation corresponds to an operation on the real system. To improve this rate, a sensible strategy seems to be to eliminate nearly insensitive coordinates by means of a parametric reduction. This approach is based on the observation that movements in these coordinates result in nearly identical feedforward signals and therefore nearly identical cost values. Thus, avoiding unnecessary exploration in these directions may yield potential benefits.

\section{Faster R2R feedforward control} \label{sec:theory}

Our proposal in this paper is to improve the convergence rate of the previously described {R2R control algorithm} by means of a dimensional reduction of the search space. The underlying idea of this approach is that some parameters may have a greater effect than others on the feedforward control law and, as a consequence, on the cost $J$ to minimize. Thus, by avoiding movements in the directions that have less impact, the algorithm should be able to find close-to-optimal solutions with fewer operations on the real system. 

In particular, we propose two different reduction methods based on the sensitivity, $\senu(t,\parnor) \in \mathbb{R}^{1\times\npar}$, of the feedforward control law~\eqref{eq:uff} to the vector $\parnor$ of control parameters. This sensitivity is given by
\begin{equation}
    \senu(t,\parnor) = \dfrac{\partial \uff(t,\parnor)}{\partial \parnor}.
\end{equation}
Note that, by computing the sensitivity with respect to these adimensional parameters, the elements of $\senu(t,\parnor)$ can be compared directly. Otherwise, if the physical parameters had different scales or magnitudes, a normalization would be mandatory. The sensitivity can be used to compute the Fisher information matrix, which is defined as
\begin{equation}
    \mathcal{F}({\parnor}) = \int_{t_0}^{\tf} \left[\senu(t,\parnor)^{\intercal}\, \senu(t,\parnor)\right] \,\dtau.
\end{equation}
Note that, by construction, the Fisher matrix $\mathcal{F}({\parnor}) \in \mathbb{R}^{\npar\times\npar}$ is symmetric and positive semidefinite.

The first reduction method consists in finding the parameters in $\parnor$ with the greatest influence on the feedforward control term. Since the sensitivity is a time-dependent signal, let us define the integral-square sensitivity, ${S_\mathrm{IS}}$, as the vector
\begin{equation}\label{eq:ISS}
S_\mathrm{IS}(\parnor)= \int_{t_0}^{\tf} \left[ \senu(\tau,\parnor) \odot \senu(\tau,\parnor)\right] \, \dtau.
\end{equation}
The components of $S_\mathrm{IS}(\parnor)\in \mathbb{R}^{1\times\npar}$ quantify the influence of each element of $\parnor$ on $\uff$. Therefore, they can be used to prioritize the parameters according to their relevance and hence to discard the least relevant ones. In this work we assume that the R2R search is conducted in the vicinity of the nominal value of the parameters, i.e., around $\parnornom=1_\npar$, where $1_\npar$ is the vector of ones of size~$\npar$. Thus, the choice of which parameters to optimize and which to keep fixed is determined by $S_\mathrm{IS}(\parnornom)$.

The above method provides a way to eliminate the least influential parameters of $\parnor$, but it is still possible that two or more parameters classified as highly relevant are somewhat correlated, i.e., have very similar influences on $\uff$. To work around this, the second reduction method is based on finding an alternative orthogonal coordinate system in which the components are ordered according to their influence on the feedforward term. For that, let us define {$D(\parnor)$} as the integral-square deviation of $\uff$ with respect to the nominal input.
\begin{equation}
    D(\parnor) = \frac{1}{2}\int_{t_0}^{\tf}\big(\uff(\tau,\parnor) -\uff(\tau,\parnornom)\big)^2 \, \dtau
\end{equation}
An approximation of $D(\parnor)$ can be obtained using the second-order Taylor expansion around the nominal parameter vector, 
\begin{equation}
    D(\parnor) \approx D(\parnornom) +  D_\parnor ({\parnornom}) \, \delta\parnor +  \frac{1}{2}\,\delta\parnor^\intercal\, D_{\parnor\parnor}({\parnornom})\,\delta\parnor,
\end{equation}
where 
\begin{equation}
    \delta\parnor =\parnor-\parnornom, \hspace{1.5em}
    D_\parnor(\parnor) = \dfrac{\partial D}{\partial \parnor}, \hspace{1.5em}
    D_{\parnor\parnor}(\parnor) = \dfrac{\partial^2 D}{\partial \parnor^2} \nonumber.
\end{equation}
It is not difficult to see that, while both $D(\parnornom)$ and $D_\parnor(\parnornom)$ are equal to zero, the Hessian matrix of $D$ at $\parnornom$ is equal to the Fisher matrix evaluated at the same point, i.e., \mbox{$D_{\parnor\parnor}(\parnornom)  = \mathcal{F}({\parnornom})$}. Therefore, the integral-square deviation of $\uff$ with respect to the nominal input is approximately given by the quadratic form
\begin{equation}
    D(\parnor) \approx \frac{1}{2}\,\delta\parnor^\intercal\, \mathcal{F}({\parnornom})\,\delta\parnor.
\end{equation}

Now, the goal is to find a linear change of variables to an orthogonal coordinate system in which each component affects $D$ independently. Since $\mathcal{F}(\parnornom)$ is symmetric and positive semidefinite, the eigendecomposition and the singular value decomposition coincide. That is, the Fisher matrix can be expressed as
\begin{equation}\label{eq:eig_F}
    \mathcal{F}({\parnornom}) = V \, \Lambda \, V^\intercal= \sum_{i=1}^{\npar} v_i\,\lambda_i\,v_i^\intercal,
\end{equation}
where $V\in \mathbb{R}^{\npar\times\npar}$ is an orthogonal matrix whose columns, $v_i \in \mathbb{R}^\npar$, are the eigenvectors (or singular vectors) of $\mathcal{F}({\parnornom})$ and $\Lambda\in \mathbb{R}^{\npar\times\npar}$ is a diagonal matrix with the corresponding eigenvalues (or singular values), $\lambda_i \in \mathbb{R}$, in the diagonal elements. Assuming that the eigenvalues are sorted by value,
\begin{equation}
    \lambda_1 \geq \lambda_2 \geq \dots \geq \lambda_{\npar} \geq 0,
\end{equation}
the Fisher matrix could be approximated by
\begin{equation}
    \mathcal{F}({\parnornom}) \approx \tilde{\mathcal{F}}(\parnornom)= \tilde V \, \tilde \Lambda \, \tilde V^\intercal =  \sum_{i=1}^{r} v_i\,\lambda_i\,v_i^\intercal,
\end{equation}
where $r$ is the order of the approximation ($r \leq q$) and $\tilde V\in \mathbb{R}^{\npar\times r}$ and $\tilde \Lambda \in \mathbb{R}^{r \times r}$ contain the first $r$ eigenvectors and eigenvalues, respectively. Using this reduced-order approximation, $D$ can be further approximated as
\begin{equation}
    D(\parnor) \approx \frac{1}{2}\,\delta\parnor^\intercal\, \tilde{\mathcal{F}}({\parnornom})\,\delta\parnor = \frac{1}{2}\,(\varphi -  \varphi^*)^\intercal \, \tilde\Lambda \, (\varphi -  \varphi^*),
\end{equation}
where $\varphi \in \mathbb{R}^r$ is the alternative orthogonal reduced-order parameter vector, given by
\begin{equation}
    \varphi = \tilde V^\intercal \, \delta \parnor +  \varphi^*= \tilde V^\intercal \, \left(\parnor-\parnornom\right) +  \varphi^*,
\end{equation}
and $\varphi^*$ is its nominal value, which can be chosen arbitrarily. By selecting $\varphi^*=\tilde V^\intercal \, \parnornom$, the change of variables simplifies into
\begin{equation}\label{eq:varphi_theta}
    \varphi = \tilde V^\intercal \, \parnor \ \Longleftrightarrow \ \parnor = \tilde V \, \varphi.
\end{equation}

The R2R search in this second method is then performed in the reduced space defined by the alternative parameterization $\varphi$. As in the first method, the search is assumed to be performed in the vicinity of the nominal parameter vector, so the change of basis, which is fixed, is based on the decomposition of $\mathcal{F}({\parnornom})$.

\section{Simulation results} \label{sec:simulation}

In this section, we implement the proposed parameter reduction techniques to obtain different parameterizations of the feedforward term. We evaluate the control strategy for each parameterization through simulated experiments. For all the experiments, the feedforward signal $\uff$ is initialized based on the nominal parameter values specified in Table~\ref{tb:param}. For the sake of brevity, we focus on the closing operation, i.e., $z_0$ and $\zf$ are set as the upper and lower position limits of the stroke of the movable core. The control process of the opening operation is completely equivalent. 

\subsection{Dimensional reduction of the search space}

\begin{figure}[t]
    \centering
	\subfloat[Original parameterization. \label{fig:du_dth_2}]	{\includegraphics{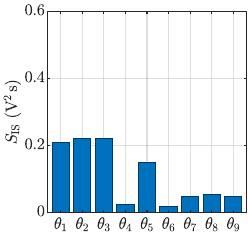}}
	\subfloat[Orthogonal parameterization.	\label{fig:du_dxi_2}]	{\includegraphics{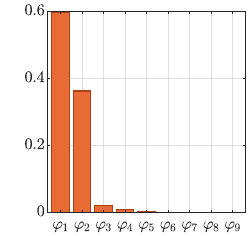}}
	\caption{Integral-square sensitivities of the feedforward control law with respect to the control parameters.}	\label{fig:du_dparam}
\end{figure}

As first step, we have computed the integral-square sensitivities, $S_\mathrm{IS}$, of the feedforward term to both the original control parameters and those resulting from orthogonalization. These sensitivities, which are displayed in Fig.~\ref{fig:du_dparam}, provide insights into potential parametric reduction cases.

\begin{table}[t]
    \begin{center}
    \renewcommand{\arraystretch}{1.1} 
    \caption{Nominal parameter values}\label{tb:param}
        \begin{tabular}{ccccc}
	    	\cmidrule{1-2} \cmidrule{4-5}
            $ \ksp $ & $ 55 \,\mathrm{N/m} $ &&
            $ \kappa_5 $ & $1320\,\mathrm{m^{-1}}$\\
            $ \zsp $ & $ 0.015 \,\mathrm{m} $ &&
            $ \kappa_6 $ & $9.73\cdot10^{-3}\,\mathrm{m}$ \\
            $ m $ & $1.6 \cdot 10^{-3} \,\mathrm{kg} $ &&
            $ R $ & $ 50 \, \mathrm{\Omega}$\\
	    	$ \auxRelcz $ & $1.35 \,\mathrm{H^{-1}} $ &&
	    	$ z_0 $ & $10^{-3}\,\mathrm{m}$ \\
	    	$ \lambdasat $ & $0.0229 \,\mathrm{Wb} $ &&
	    	$ z_\mathrm{f} $ & $0$ \\
    		$ \auxRelgz  $ & $3.88 \,\mathrm{H^{-1}} $ &&
    		$ t_0 $ & $ 0 $ \\
            $ \auxdRelgz $ & $7.67  \cdot 10^{4} \,\mathrm{H^{-1}/m} $ &&
    		$ \tf $ & $ 3.5 \cdot 10^{-3}\,\mathrm{s}$ \\	
	    	\cmidrule{1-2} \cmidrule{4-5} 
        \end{tabular}
    \end{center}
\end{table}

\begin{table}[t]
    \begin{center}
    \setlength{\tabcolsep}{3pt} 
    \renewcommand{\arraystretch}{1.1} 
    \caption{Reduction cases}\label{tb:red}
        \begin{tabular}{cccc}
            Case & \;\;\,$r$\;\;\, & Free parameters & Fixed parameters  \\
	    	\cmidrule{1-4}
	    	A & $ 9 $ & $\theta_1, \ldots, \; \theta_9$ \\
            \cmidrule{1-4}
            B & $ 7 $ & 
            $ \theta_1, \; \theta_2, \; \theta_3, \; \theta_5, \; \theta_7, \; \theta_8, \; \theta_9  $ &
            $ \theta_4, \; \theta_6 $ \\
            C & $ 4 $ & 
            $ \theta_1, \; \theta_2, \; \theta_3, \; \theta_5  $ &
            $ \theta_4, \; \theta_6, \; \theta_7, \; \theta_8, \; \theta_9 $ \\
            D & $ 2 $ & 
            $ \theta_2, \; \theta_3  $ &
            $ \theta_1, \; \theta_4, \; \theta_5, \; \theta_6, \; \theta_7, \; \theta_8, \; \theta_9 $ \\
            \cmidrule{1-4}
            E & $ 7 $ & 
            $ \varphi_1, \; \ldots, \; \varphi_7  $ &
            $ \varphi_8, \; \varphi_9 $ \\
            F & $ 4 $ & 
            $ \varphi_1, \; \ldots, \; \varphi_4 $ &
            $ \varphi_5, \; \ldots, \; \varphi_9 $ \\
            G & $ 2 $ & 
            $ \varphi_1, \; \varphi_2 $ &
            $ \varphi_3, \ldots, \; \varphi_9 $ \\
	    	\cmidrule{1-4} 
        \end{tabular}
    \end{center}
\end{table}

As previously presented, the first reduction technique requires computing the sensitivity with respect to each original parameter $\theta_i$ (the subscript $i$ denotes the $i$th vector component). These sensitivities, which are visualized in Fig.~\ref{fig:du_dth_2}, indicate two main possible reduction cases: eliminate from the search only $\theta_4$ and $\theta_6$, which are by far the least influential parameters; or eliminate $\theta_7$, $\theta_8$, and $\theta_9$ as well, because their $S_\mathrm{IS}$ values are also relatively small.

The second reduction technique involves a change of basis resulting in a new parameterization based on the orthogonal parameters. The sensitivities to these new parameters have been computed and represented in Fig.~\ref{fig:du_dxi_2}. Note that these sensitivity values, which are the eigenvalues of the Fisher matrix $\mathcal{F}({\parnornom})$, have been represented in order from the most to the least influential in the feedforward control law. Based on the visualized values, a new possible reduction case is suggested: eliminate from the search all but $\varphi_1$ and $\varphi_2$, as they are the most influential ones by a wide margin.

In summary, we have identified three promising reduced-order parameter vectors, with dimensions $r=7$ and $4$ for the original parameters, and $r=2$ for the orthogonal parameters. Nonetheless, in order to ensure a comprehensive comparison, in the following we evaluate equivalent parametric reductions, with $r=7$, $4$, and $2$, to both sets of parameters. We also consider a baseline scenario in which no reduction is performed, i.e., $r=q=9$. All studied reduction cases are summarized in Table~\ref{tb:red}.

\subsection{Control results}

The controller performance with the proposed reductions has been evaluated through simulations using the dynamical model presented in Section~\ref{sec:model}. We assume that the controlled dynamic system is governed by these equations. However, considering that the parameters are never perfectly known in practice, it is assumed that there is some uncertainty in their values. In particular, each component of the model parameter vector $p$ is randomly and independently perturbed between $95\,\%$ and $105\,\%$ of its nominal value (see Table~\ref{tb:param}). Note that the electrical resistance $R$ is not perturbed because it is typically measured with high precision. For each reduction case, 10\,000 different trials have been simulated, and in each trial the control algorithm is run for 300 switching operations. The feedforward control law, on the other hand, starts all trials using the nominal value, i.e., $\parnor^{[1]}=\parnornom=1_\npar$.

\begin{figure}[p]
    \subfloat[Case A: No reduction ($r=9$).	\label{fig:vc_th9}]	{\includegraphics[trim={0 0 0 0.01mm},clip]{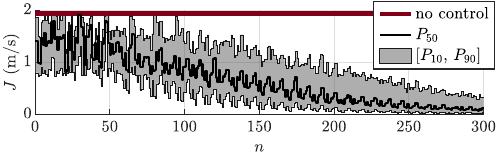}}\vcap
    \subfloat[Case B: Original parameters ($r=7$).	\label{fig:vc_th7}]{\includegraphics[trim={0 0 0 0.01mm},clip]{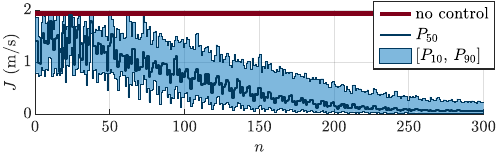}}\vcap
    \subfloat[Case C: Original parameters ($r=4$).	\label{fig:vc_th4}]	{\includegraphics[trim={0 0 0 0.01mm},clip]{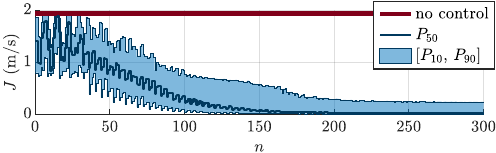}}\vcap
    \subfloat[Case D: Original parameters ($r=2$).	\label{fig:vc_th2}]	{\includegraphics[trim={0 0 0 0.01mm},clip]{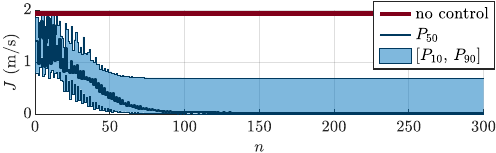}}\vcap
	\subfloat[Case E: Orthogonal parameters ($r=7$). \label{fig:vc_xi7}] {\includegraphics[trim={0 0 0 0.01mm},clip]{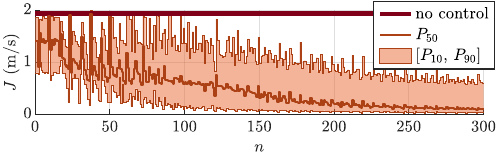}}\vcap
	\subfloat[Case F: Orthogonal parameters ($r=4$). \label{fig:vc_xi4}] {\includegraphics[trim={0 0 0 0.01mm},clip]{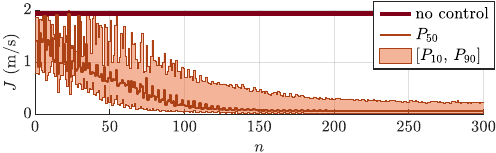}}\vcap
	\subfloat[Case G: Orthogonal parameters ($r=2$). \label{fig:vc_xi2}] {\includegraphics[trim={0 0 0 0.01mm},clip]{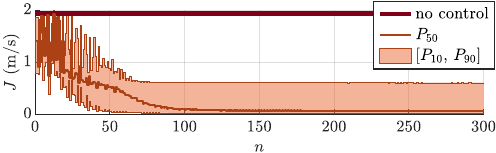}}\vcap
	\caption{Cost values with respect to the number of switching operations. Each graph shows the median ($P_{50}$) and the 10th and 90th percentiles ($P_{10}$ and $P_{90}$, respectively) of the distribution of values obtained for the 10\,000 simulated experiments. The cost without control is also represented.}	\label{fig:vc}
\end{figure}

The control results for each case are summarized in Fig.~\ref{fig:vc}, where each graph represents the obtained distribution of costs, $J = \lvert v_\mathrm{c}\rvert$, with respect to the switching operation, $n$. To show the effectiveness of each control scenario, the graphs also display the cost of a conventional switching operation, namely with a 30~V constant activation.

For the first simulated case, serving as a baseline, no parameter reduction is applied (case A from Table~\ref{tb:red}). Note that this scenario corresponds to the direct application of the control algorithm presented in~\cite{moya2023IFAC}. As shown in the control results (see Fig.~\ref{fig:vc_th9}), the disparity between the system parameters (randomly perturbed) and the initial feedforward parameters results in a large variability in the impact velocities of the first switching operation ($n=1$). Then, due to the adaptation law, the control performance improves as the number of iterations increases. However, it is evident that the control convergence is quite slow, as it requires more than 300 operations to reach stable values of the cost. The main reason for this is the large number of parameters modified by the adaptation law.

Subsequently, the parameter reductions are applied with the aim of improving the controller convergence. In the next simulated cases, the reduced original parameter vectors (with dimensions $r=7$, $4$, and $2$) are used for the adaptation law. The control results are displayed in Figs.~\ref{fig:vc_th7}, \ref{fig:vc_th4}, and~\ref{fig:vc_th2}, respectively. As can be seen, the number of iterations required to improve the performance is reduced. In particular, with no parameter reduction, the control needs to perform 203 operations to halve the cost of the uncontrolled scenario in 90\,\% of the trials. In contrast, with this type of parameter reduction, the control strategy only requires 153, 83, and 42 operations in each subsequent case.

Furthermore, the results with the reduced orthogonal parameters are visualized in Figs.~\ref{fig:vc_xi7}, \ref{fig:vc_xi4}, and~\ref{fig:vc_xi2} for $r=7$, $4$, and $2$, respectively. These also demonstrate a significant improvement in the control convergence rate with respect to the baseline. However, when each case is compared with its equivalent case (i.e., with the same $r$) using the original parameterization, a significant similarity is observed, making it challenging to discern the superior performer. For instance, concerning $r=7$, case E (with the orthogonal parameterization) presents better results on average than case B (using the original parameters). However, it also exhibits a higher variability. This suggests that the poor performance of case E may be due to the negligible effect of the last included orthogonal parameters, $\varphi_5$ to $\varphi_7$, on the feedforward control law. In contrast, for $r=4$ and $r=7$ the results with both reduction techniques are very similar.

\begin{figure}[t]
    \centering
    \includegraphics[trim={0 1mm 0 0},clip]{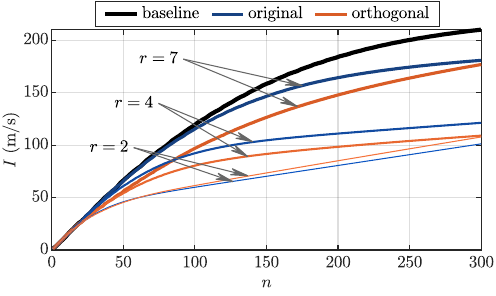}
    \caption{Integrated costs with respect to the number of iterations.  Each line represents the mean values for the 10\,000 simulated experiments. Comparison between different reduction methods.}
    \label{fig:vc_cums_742}
\end{figure}

To facilitate comparisons across all the studied cases, we have computed the integrated (i.e., cumulative) cost of each trial, denoted as $I$, which provides a measure of the overall performance of each control strategy for any iteration count.
\begin{equation}
    I^{[n]} = \sum_{i=1}^{n} J^{[n]}
\end{equation}
The average values of these integrated costs across all $10\,000$ trials are represented in Fig.~\ref{fig:vc_cums_742}. For any count of the tested operations, from 1 to 300, it is evident that reducing the dimension $r$ of the search space improves the total cost. This representation also shows that reductions with orthogonal parameters have superior performance for $r=7$ and $r=4$, although it is slightly less effective for $r=2$. Moreover, by examining the $I$ values and their slopes for the last operation ($n=300$), the main drawback of extreme parametric reduction ($r=2$) becomes apparent: if the number of operations were to increase further, the total control cost would grow faster than in the other cases, because these have converged to worse stable values. These results also demonstrate that, for a larger number of iterations, the best performance is achieved with the orthogonal parameterization of dimension $r=4$. In any case, the results show that the control convergence is improved in all the considered scenarios, which emphasizes the potential of the proposed sensitivity-based reduction methodology.
\FloatBarrier

\section{Conclusions}
\label{sec:conclusions}

In this work, we have analyzed the possibility of improving the convergence rate of an adaptive feedforward R2R control algorithm by means of a sensitivity-based dimensional reduction of the search space. The two strategies presented show great potential in simulation, but the results obtained do not yet allow us to determine whether one of the two is more efficient than the other. Possibly, this will depend on the system to be controlled and the physical parameterization originally chosen for the dynamical model.

As future work, we would like to evaluate the possibility of periodically updating the reduced parametric basis. That is, the two proposals of this work use a fixed transformation based on the nominal value of the parameters, but it would be interesting to evaluate whether it is advantageous to redo the calculations every iteration or every few iterations. In addition, we also intend to perform real laboratory tests and verify that the experimental results agree with those observed in simulation.


\begin{thebibliography}{10}
\providecommand{\url}[1]{#1}
\csname url@rmstyle\endcsname
\providecommand{\newblock}{\relax}
\providecommand{\bibinfo}[2]{#2}
\providecommand\BIBentrySTDinterwordspacing{\spaceskip=0pt\relax}
\providecommand\BIBentryALTinterwordstretchfactor{4}
\providecommand\BIBentryALTinterwordspacing{\spaceskip=\fontdimen2\font plus
\BIBentryALTinterwordstretchfactor\fontdimen3\font minus \fontdimen4\font\relax}
\providecommand\BIBforeignlanguage[2]{{%
\expandafter\ifx\csname l@#1\endcsname\relax
\typeout{** WARNING: IEEEtran.bst: No hyphenation pattern has been}%
\typeout{** loaded for the language `#1'. Using the pattern for}%
\typeout{** the default language instead.}%
\else
\language=\csname l@#1\endcsname
\fi
#2}}

\bibitem{van2019design}
J.~van Dam, D.~Kropl, B.~Gysen, and E.~Lomonova, ``Design of a permanent magnet-biased reluctance valve actuator with integrated eddy current damping,'' in \emph{Int. Symp. Linear Drives for Ind. Applic.}\hskip 1em plus 0.5em minus 0.4em\relax IEEE, 2019, pp. 1--6.

\bibitem{DiGaeta2015}
A.~Di~Gaeta, C.~I. Hoyos~Velasco, and U.~Montanaro, ``Cycle-by-cycle adaptive force compensation for the soft-landing control of an electro-mechanical engine valve actuator,'' \emph{Asian J. Control}, vol.~17, no.~5, pp. 1707--1724, Jul. 2015.

\bibitem{bojan2019design}
B.~Gergič and D.~Hercog, ``Design and implementation of a measurement system for high-speed testing of electromechanical relays,'' \emph{Measurement}, vol. 135, pp. 112--121, Mar. 2019.

\bibitem{deschaux2019magnetic}
F.~Deschaux, F.~Gouaisbaut, and Y.~Ariba, ``Magnetic force modelling and nonlinear switched control of an electromagnetic actuator,'' in \emph{IEEE Conf. Decision and Control}.\hskip 1em plus 0.5em minus 0.4em\relax IEEE, 2019, pp. 1416--1421.

\bibitem{fang2019modeling}
J.~Fang, X.~Wang, J.~Wu, S.~Yang, L.~Li, X.~Gao, and Y.~Tian, ``Modeling and control of a high speed on/off valve actuator,'' \emph{Int. J. Automot. Technol.}, vol.~20, pp. 1221--1236, 2019.

\bibitem{Benosman2015}
M.~Benosman and G.~M. At{\i}n{\c{c}}, ``{Extremum seeking-based adaptive control for electromagnetic actuators},'' \emph{Int. J. Control}, vol.~88, no.~3, pp. 517--530, 2015.

\bibitem{hoffmann2003iterative}
W.~Hoffmann, K.~Peterson, and A.~G. Stefanopoulou, ``Iterative learning control for soft landing of electromechanical valve actuator in camless engines,'' \emph{IEEE Trans. Control Syst. Technol.}, vol.~11, no.~2, pp. 174--184, Mar. 2003.

\bibitem{braun2019flatness}
T.~Braun, J.~Reuter, and J.~Rudolph, ``Flatness-based feed-forward control design for solenoid actuators considering eddy currents,'' \emph{IFAC-PapersOnLine}, vol.~52, no.~15, pp. 567--572, 2019, 8th IFAC Symp. Mechatronic Syst.

\bibitem{ramirez2017new}
E.~Ramirez-Laboreo, C.~Sagues, and S.~Llorente, ``A new run-to-run approach for reducing contact bounce in electromagnetic switches,'' \emph{IEEE Trans. Ind. Electron.}, vol.~64, no.~1, pp. 535--543, Jan. 2017.

\bibitem{serrano2022}
E.~Serrano-Seco, E.~Ramirez-Laboreo, E.~Moya-Lasheras, and C.~Sagues, ``An audio-based iterative controller for soft landing of electromechanical relays,'' \emph{IEEE Trans. Ind. Electron.}, vol.~70, no.~12, pp. 12\,730--12\,738, Dec. 2023.

\bibitem{lewis2000pattern}
R.~M. Lewis and V.~Torczon, ``Pattern search methods for linearly constrained minimization,'' \emph{SIAM J. Optimization}, vol.~10, no.~3, pp. 917--941, 2000.

\bibitem{moya2019novel}
E.~Moya-Lasheras, E.~Ramirez-Laboreo, and C.~Sagues, ``{A novel algorithm based on Bayesian optimization for run-to-run control of short-stroke reluctance actuators},'' in \emph{European Control Conf.}\hskip 1em plus 0.5em minus 0.4em\relax IEEE, Jun. 2019, pp. 1103--1109.

\bibitem{miao2011identifiability}
H.~Miao, X.~Xia, A.~S. Perelson, and H.~Wu, ``On identifiability of nonlinear ode models and applications in viral dynamics,'' \emph{SIAM review}, vol.~53, no.~1, pp. 3--39, 2011.

\bibitem{walter1997identification}
E.~Walter, L.~Pronzato, and J.~Norton, \emph{Identification of parametric models from experimental data}.\hskip 1em plus 0.5em minus 0.4em\relax Springer, 1997, vol.~1, no.~2.

\bibitem{moya2023IFAC}
E.~Moya-Lasheras, E.~Ramirez-Laboreo, and E.~Serrano-Seco, ``{Run-to-Run Adaptive Nonlinear Feedforward Control of Electromechanical Switching Devices},'' \emph{IFAC-PapersOnLine}, vol.~56, no.~2, pp. 5358--5363, 2023, 22nd IFAC World Congr.

\end{thebibliography}
\end{document}